\documentclass[journal]{IEEEtran}
\usepackage{graphicx}
\usepackage{float}
\usepackage{amsthm}
\usepackage{amssymb}
\usepackage{textcomp}
\usepackage{nomencl}
\makenomenclature
\usepackage{ifthen}
\usepackage{setspace}
\usepackage{multirow}
\usepackage{paralist}

\usepackage{cite}
\usepackage{amsmath}
\usepackage{amsmath,bm}
\usepackage{algorithmic}
\usepackage{algorithm}

\usepackage{array}

\ifCLASSOPTIONcompsoc
  \usepackage[caption=false,font=normalsize,labelfont=sf,textfont=sf]{subfig}
\else
  \usepackage[caption=false,font=footnotesize]{subfig}
\fi
\newtheorem{proposition}{Proposition}
\newtheorem{defination}{Definition}

\begin{document}
%
\title{A Global Solution Method for Decentralized Multi-Area SCUC and Savings Allocation Based on MILP Value Functions}



\author{Xiaodong~Zheng,
        Haoyong~Chen,
        Yan Xu,
        Feifan Shen,
        Zipeng Liang
\thanks{This work was supported by National Key Research and Development Program of China under grant 2016YFB0900100.} 
\thanks{X. Zheng, H. Chen and Z. Liang are with the School of Electric Power, South China University of Technology, Guangzhou 510641, China. X. Zheng is also a visiting Ph.D. student in the School of Electrical and Electronic Engineering, Nanyang Technological University, Singapore (e-mail: z.xiaodong@mail.scut.edu.cn, eehychen@scut.edu.cn, liangzipeng.ye@163.com).}
\thanks{Y. Xu is with the School of Electrical and Electronic Engineering, Nanyang Technological University, Singapore (e-mail: eeyanxu@gmail.com).} 
\thanks{F. Shen is with the Centre for Electric Power and Energy, Department of Electrical Engineering, Technical University of Denmark, Kgs. Lyngby, DK 2800 (e-mail: fshen@elektro.dtu.dk).}}

\markboth{ }%
{Shell \MakeLowercase{\textit{et al.}}: Bare Demo of IEEEtran.cls for IEEE Journals}
%



\IEEEtitleabstractindextext{
\begin{abstract}
To address the issue that Lagrangian dual function based algorithms cannot guarantee convergence and global optimality for decentralized multi-area security constrained unit commitment (M-SCUC) problems, a novel decomposition and coordination method using MILP (mixed integer linear programming) value functions is proposed in this paper. Each regional system operator sets the tie-line power injections as variational parameters in its regional SCUC model, and utilizes a finite algorithm to generate a MILP value function, which returns the optimal generation cost for any given interchange scheduling. With the value functions available from all system operators, theoretically, a coordinator is able to derive a globally optimal interchange scheduling. Since power exchanges may alter the financial position of each area considerably from what it would have been via scheduling independently, we then propose a fair savings allocation method using the values functions derived above and the Shapley value in cooperative game theory. Numerical experiments on a two-area 12-bus system and a three-area 457-bus system are carried out. The validness of the value functions based method is verified for the decentralized M-SCUC problems. The outcome of savings allocation is compared with that of the locational marginal cost based method.
\end{abstract}

}

\maketitle

\IEEEdisplaynontitleabstractindextext

%
\IEEEpeerreviewmaketitle



\section{Introduction}\label{sec:introduction}
%
%
%
%
\IEEEPARstart{T}{he} real-world power systems are usually composed of several regional sub-systems interconnected by tie-lines. Such bulk power systems are referred to as multi-area power systems. In the context of short-term operation, multi-area security constrained unit commitment (M-SCUC) models have been widely applied in the market clearing or generation/interchange scheduling of multi-area power systems~\cite{shoults1980practical, kargarian2014distributed, li2015decentralized}. One of the crucial issues of M-SCUC is the interchange scheduling, which can save considerable operating costs if optimally performed.

The essential difference between M-SCUC and single-area SCUC problems is that under most circumstances, M-SCUC should be performed in a decentralized manner. This is due to the way how regional system operators and the higher-level system operator (or the coordinator) are organized, e.g., the hierarchical control architecture in China~\cite{li2015decentralized-ed, zheng2017loss}, and the proxy bus systems in the US~\cite{ji2017multi, guo2017coordinated}. Though easy for implementation, current methods may suffer from sub-optimality or market inefficiency~\cite{zheng2017loss, guo2017coordinated}.

The challenge of solving M-SCUC is to achieve \emph{joint optimality} of this complex problem and meanwhile preserve the decentralized decision-making procedure among regional operators. Theoretically, this can be tackled by solving a decentralized mathematical programming problem. For convex power system optimal dispatch problems, efficient decentralized solution methods can be easily found~\cite{molzahn2017survey}. Among these methods, augmented Lagrangian function based methods, e.g., alternating direction method of multipliers (ADMM), auxiliary problem principle (APP) are most popular~\cite{erseghe2014distributed, zheng2015fully}. Besides, other methods like modified generalized benders decomposition method~\cite{li2015decentralized-ed}, cutting plane consensus method~\cite{zhao2016fully}, etc. are applicable as well.

However, the above methods are developed based on the convexity condition of the optimization problem. For M-SCUC, due to the non-convexity caused by integer variables, strong duality doesn't hold, and hence these methods cannot guarantee convergence and global optimality. Heuristics or parameters tunings are needed to ensure the algorithms converge to a local optimum~\cite{li2015decentralized, feizollahi2015large, ahmadi2013multi}.

In the community of operations research, some contributions have been made recently on the topic of solving decentralized mixed integer linear programming (MILP) problems. For example, the authors in~\cite{vujanic2016decomposition} verify that the coupling constraints of the primal MILPs can be modified to guarantee that the solutions recovered from the dual are feasible, and the quality of solutions also improves as the size of coupling constraints increases. However, the coupling constraints in M-SCUC (i.e., consensus conditions of tie-line power or boundary phase angle) are ``hard'' conditions, which cannot be relaxed. In~\cite{feizollahi2017exact}, it is shown that the duality gap of some types of augmented Lagrangian dual can be closed, but the resulting sub-problems are always inseparable. In~\cite{testa2018distributed}, the authors propose a fully distributed algorithm for MILP. In the proposed scheme, each agent solves a linear programming (LP) and exchanges active constraints to update the LP model, and the algorithm will converge a suboptimal solution up to any given tolerance. More recently, reference~\cite{boland2018parallelizable} proposes a parallelizable augmented Lagrangian method, which is proved to converge to the optimal solution under mild conditions. But the method relies on a subroutine of refining the convex hulls of the feasible regions of each sub-problems, which may be intractable for large-scale problems.

On the other hand, after the power interchange is scheduled, conventionally the locational marginal prices (LMPs) derived from the the Lagrangian multipliers of the SCUC problem are adopted for market clearing or cost allocation. However, it is well-known that the Lagrangian multiplier of a MILP cannot depict the real change in optimal cost as the right-hand side is perturbed, due to its non-convexity. Thus the price signal is unnecessary true. Actually, it is vital to \emph{fairly} allocate the resulting economic benefits of coordinately interchange scheduling~\cite{rau2005inter}. For a non-convex market, we can turn to the cooperative game theory (e.g., the Shapley value~\cite{chapman2017cooperative}) for a promising method of cost or profit allocation~\cite{o2009nonconvex, hu2006allocation}.

We find that another type of dual function, i.e., the \emph{value function}~\cite{hassanzadeh2014on}, naturally provides a zero-gap with respect to the primal problem, and therefore can be used to solve the decentralized M-SCUC problem. Moreover, it is applicable for savings allocation when we use the Shapley value. In mathematical programming, the value function is a real-valued function that takes a right-hand side vector as input, and returns the optimal objective value of the programming instance, which is parameterized on that right-hand side vector~\cite{hassanzadeh2014on}. The main contributions of this paper are summarized below.
\begin{enumerate}
  \item We study the MILP value function of the SCUC problem, and develop a finite algorithm for its construction.
  \item Based on the SCUC value functions, we then propose a new global solution method for the decentralized M-SCUC problem, which needs limited information sharing and no iterations between system operators.
  \item Using the SCUC value functions, we derive a fair and stable savings allocation scheme based on the Shapley value, and compare it with the LMPs based method.
  \item The effectiveness of our method is demonstrated on a two-area 12-bus system and a three-area 457-bus system.
\end{enumerate}

\section{Preliminaries}\label{sec:preliminary}
\begin{figure}
  \centering
  \includegraphics[width=3.3in]{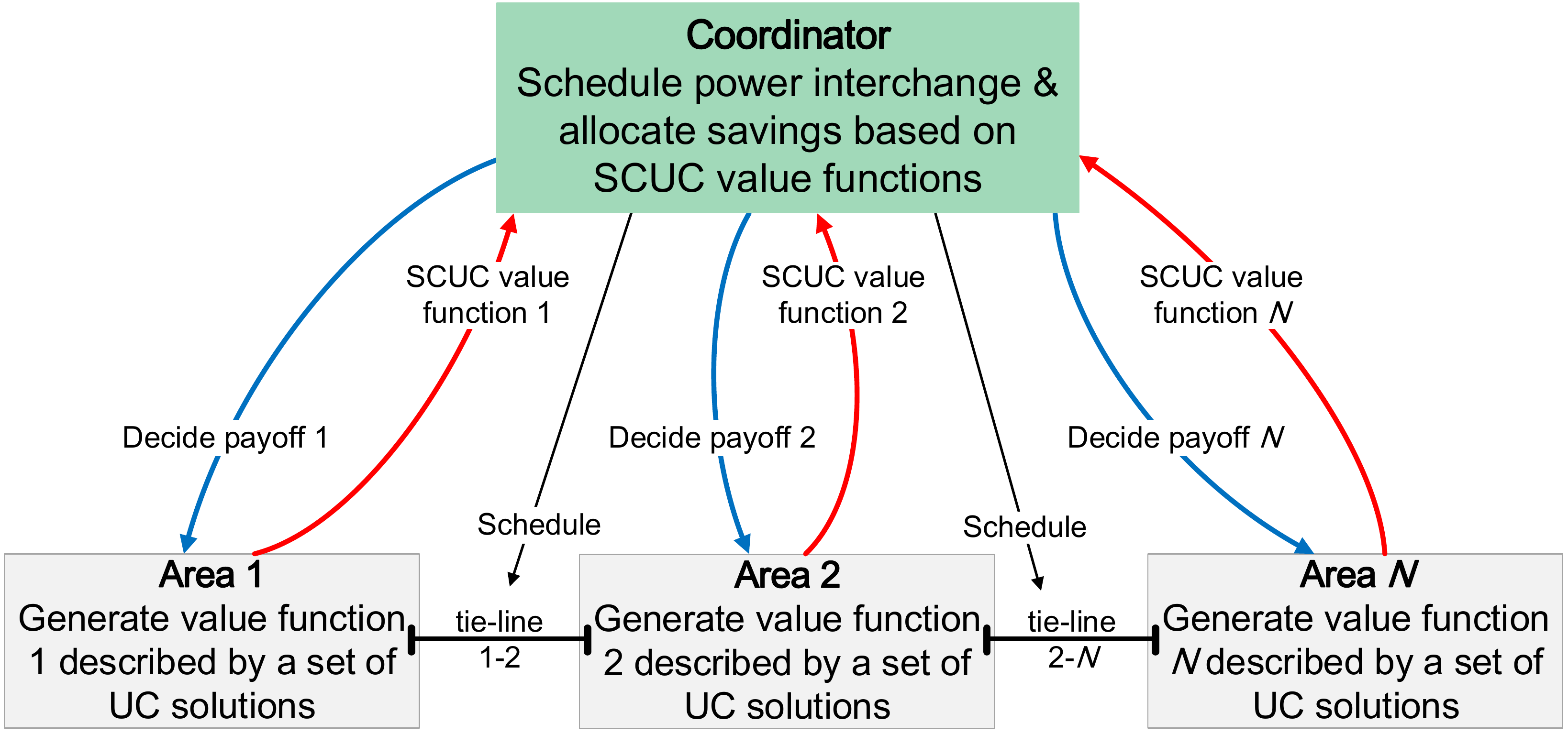}\\
  \caption{Framework of the decentralized solution procedure.}\label{fig:Framework}
\end{figure}

The framework of the proposed method is given in Fig.~\ref{fig:Framework}. In order to solve the M-SCUC problem in a decentralized manner, the solution procedure is divided into two steps. First, each regional system generates its value function described by a set of UC solutions; then, the coordinator collects the value functions to determine the interchange scheduling, and meanwhile decide the systems' payoffs. In this section, we present the M-SCUC model, and introduce the concepts of MILP value function and Shapley value first.

\subsection{Multi-area Power Systems SCUC Problem}\label{sec:preliminary_M-SCUC}
Conventionally, the SCUC problem is modeled as a MILP~\cite{chen2016key}. For the M-SCUC model, we choose tie-line injections as the coupling variables between areas. The full formulation is removed temporarily due to space limitations. Hereafter, we focus on its compact matrix formulation,
\begin{subequations}\label{eqn:M-SCUC}
    \begin{align}
        \underset{\bm{x}_{a},\bm{y}_{a},\bm{z}_{a}}{\mathop{\min }}\,& \sum\limits_{a\in \mathsf{\mathcal{A}}}{{\bm{c}_{Ia}^{\top}}{\bm{x}}_{a}}+{\bm{c}_{Ca}^{\top}} {\bm{y}}_{a}~~~~~~ \label{eqn:M-SCUC_obj}\\
        \operatorname{s.t.~} & {{\bm{G}}_{a}}{{\bm{x}}_{a}} \le {{\bm{g}}_{a}} ~~ \forall a\in \mathsf{\mathcal{A}}  \label{eqn:M-SCUC_Gg}\\
        & {{\bm{H}}_{a}}{{\bm{y}}_{a}}\le {{\bm{h}}_{a}} ~~ \forall a\in \mathsf{\mathcal{A}}  \label{eqn:M-SCUC_Hh}\\
        & {{\bm{A}}_{a}}{{\bm{x}}_{a}}+{{\bm{D}}_{a}}{{\bm{y}}_{a}}\le \bm{0} ~~ \forall a\in \mathsf{\mathcal{A}}  \label{eqn:M-SCUC_AD}\\
        & {{\bm{B}}_{a}}{{\bm{y}}_{a}} + {{\bm{z}}_{a}} = \bm{d}_{a} ~~ \forall a\in \mathsf{\mathcal{A}}  \label{eqn:M-SCUC_B}\\
        & {{\bm{z}}_{a}^{\min}} \le {{\bm{z}}_{a}} \le {{\bm{z}}_{a}^{\max}} ~~ \forall a\in \mathsf{\mathcal{A}} \label{eqn:M-SCUC_z-range} \\
        & \sum\limits_{a\in \mathsf{\mathcal{A}}}{{{\bm{C}}_{a}}}{{\bm{z}}_{a}}=\bm{0}, \label{eqn:M-SCUC_consensus}
    \end{align}
\end{subequations}
where $\mathcal{A}$ is the set of area, $\bm{x}_a$ indicates the 0-1 binary unit commitment variables of area $a$, $\bm{y}_a$ the continuous dispatch variables of area $a$, $\bm{z}_a$ the tie-line injection variables of area $a$, and $\bm{c}_{Ia}$, $\bm{c}_{Ca}$ are the cost coefficients associated with $\bm{x}_a$ and $\bm{y}_a$ respectively. Constraint (\ref{eqn:M-SCUC_Gg}) includes the state transition equations of units and minimum on/off time limits of units; constraint (\ref{eqn:M-SCUC_Hh}) includes ramping limits of units, power flow limits on internal transmission lines; constraint (\ref{eqn:M-SCUC_AD}) is the coupling constraint of $\bm{x}_a$ and $\bm{y}_a$ (i.e., the generation capacity limit); constraint (\ref{eqn:M-SCUC_B}) represents the power balance condition (e.g., the DC power flow equation); constraint (\ref{eqn:M-SCUC_z-range}) represents the power flow limits of tie-lines; and constraint (\ref{eqn:M-SCUC_consensus}) is the consensus condition between areas, indicating that the nodal injections of two ends of a tie-line should be identical. The right-hand side vector $\bm{g}_a$ is composed of the minimum on/off time of units, $\bm{h}_a$ the ramping capacities of units, transmission line ratings, reserve requirements, etc., and $\bm{d}_a$ the nodal injections (load demands).

Problem (\ref{eqn:M-SCUC}) can be written as a two-stage optimization problem, i.e.,
\begin{subequations}\label{eqn:M-SCUC_new}
    \begin{align}
        \underset{\bm{z}_{a}}{\mathop{\min }}\,& \sum\limits_{a\in \mathsf{\mathcal{A}}}{V_a(\bm{z}_{a})} \label{eqn:M-SCUC_obj_new}\\
        \operatorname{s.t.~} & \text{(\ref{eqn:M-SCUC_z-range})},\text{(\ref{eqn:M-SCUC_consensus})} \\
        & V_a(\bm{z}_{a}) = \underset{\bm{x}_{a},\bm{y}_{a}}{\mathop{\min }}\,{{\bm{c}_{Ia}^{\top}}{\bm{x}}_{a}}+{\bm{c}_{Ca}^{\top}} {\bm{y}}_{a} \\
        & ~~~~~~~~~~~~~~~~~~\operatorname{s.t.~} \text{(\ref{eqn:M-SCUC_Gg})},\text{(\ref{eqn:M-SCUC_Hh})},\text{(\ref{eqn:M-SCUC_AD})},\text{(\ref{eqn:M-SCUC_B})}.
    \end{align}
\end{subequations}
Herein, $V_a(\bm{z}_{a})$ is a function parameterized on the tie-line injection vector $\bm{z}_{a}$, and it stands for the SCUC problem of area $a$. By eliminating free variables $\bm{y}_a$ and inequality constraints (\ref{eqn:M-SCUC_Gg})-(\ref{eqn:M-SCUC_AD}) with nonnegative slack variables, the $a$-th SCUC problem can be reduced to a standard MILP as below~\cite{bertsimas1997introduction},
\begin{equation}
V_a(\bm{z}_{a})=\underset{(\bm{x_a},\bm{y_a})\in S_a(\bm{z}_{a})}\min {\bm{c}}_{Ia}^{\top}{\bm{x}_{a}} + {\bm{c}}_{Ca}^{\top}{\bm{y}_{a}},
\label{eq:Standard_MILP}
\end{equation}
where $S_a(\bm{z}_a)$ is the feasible region, i.e.,
\begin{equation} \label{eqn:S_z}
S_a(\bm{z}_a) = \left\{ (\bm{x},\bm{y})\in \mathbb{B}^{m_a}\times \mathbb{R}_{+}^{n_a}:{{\bm{A}}_{Ia}}\bm{x}_a+{{\bm{A}}_{Ca}}\bm{y}_a=\tilde{\bm{z}}_a \right\},
\end{equation}
and $\mathbb{B}$ denotes $\{0, 1\}$, $\mathbb{R}_{+}$ denotes $[0,+\infty)$.
In the standard MILP (\ref{eq:Standard_MILP}), decision variables $\bm{x}_a$ and $\bm{y}_a$ are augmented with slack variables, parameter matrix ${{\bm{A}}_{Ia}}$ and ${{\bm{A}}_{Ca}}$ are composed of the data associated with the binary variables and continuous variables in problem (\ref{eqn:M-SCUC}) respectively, and the right-hand side vector $\tilde{\bm{z}}_a$ is the combination of $\bm{z}_a$, $\bm{g}_a$, $\bm{h}_a$ and $\bm{d}_a$, etc.

\subsection{Value Function of SCUC}\label{sec:preliminary_VF}
In engineering practice, it is helpful to realize how the optimal solution or the optimal objective value would change, when the parameter of an optimization problem is slightly perturbed. This is known as the \emph{sensitivity analysis}. Accordingly, determining how the optimal objective value varies with the parameters in a full range falls into the scope of \emph{parametric programming}~\cite{wittmann2013global, guo2017coordinated}, or the study of \emph{value function}~\cite{hassanzadeh2014on}.

In this subsection, we take the SCUC problem (\ref{eq:Standard_MILP}) for example to illustrate the concept of MILP value function. For brevity, herein the subscript $a$ is omitted and $\tilde{\bm{z}}$ is simply denoted as $\bm{z}$. Thus, set (\ref{eqn:S_z}) can be rewritten as,
\begin{equation} \label{eqn:S_z_new}
S(\bm{z}) = \left\{ (\bm{x},\bm{y})\in \mathbb{B}^{m}\times \mathbb{R}_{+}^{n}:{{\bm{A}}_{I}}\bm{x}+{{\bm{A}}_{C}}\bm{y}=\bm{z} \right\}.
\end{equation}

\begin{defination} \label{def:VF}
The value function of MILP is a real-valued function $V:{{\mathbb{R}}^{r}}\to \mathbb{R}$ parameterized on $\bm{z}$, i.e.,
\begin{equation}\label{eq:MILP_VF_def}
V(\bm{z})=\underset{\left( \bm{x},\bm{y} \right)\in S(\bm{z})}{\mathop{\min }}\,\bm{c}_{I}^{\top}\bm{x}+\bm{c}_{C}^{\top}\bm{y}~~~\forall \bm{z}\in Z.
\end{equation}
where $Z=\left\{ \bm{z}\in {{\mathbb{R}}^{r}}:S\left( \bm{z} \right)\ne \varnothing  \right\}$ is the domain of $\bm{z}$ within which the minimum of (\ref{eq:MILP_VF_def}) is attainable.
\end{defination}

Besides, we let
\begin{align*}
& {{S}_{C}}( \bm{x},\bm{z} )=\left\{ \bm{y}\in \mathbb{R}_{+}^{n}:\allowbreak{{\bm{A}}_{I}}\bm{x}+{{\bm{A}}_{C}}\bm{y}=\bm{z} \right\}, \\
& {{S}_{I}}( \bm{z} )=\left\{ \bm{x}\in \mathbb{B}^{m}:\allowbreak{{S}_{C}}\left( \bm{x},\bm{z} \right)\ne \varnothing  \right\}, \\
& \text{and~}{{S}_{I}} = {{\cup }_{\bm{z}\in Z}}{{S}_{I}}\left( \bm{z} \right).
\end{align*}
By construction, ${{S}_{I}}$ is the set of all binary solutions that making set ${{S}_{C}}( \bm{x},\bm{z} )$ unempty. Then, for all $\bm{z}\in Z$ we have
\begin{subequations}
    \begin{align}
        V(\bm{z})& =\underset{\left( \bm{x},\bm{y} \right)\in S(\bm{z})}{\mathop{\min }}\,\bm{c}_{I}^{\top}\bm{x}+\bm{c}_{C}^{\top}\bm{y} \\
        & =\underset{\bm{x}\in {{S}_{I}}\left( \bm{z} \right)}{\mathop{\min }}\,\underset{\bm{y}\in {{S}_{C}}(\bm{x},\bm{z})}{\mathop{\min }}\,\bm{c}_{I}^{\top}\bm{x}+\bm{c}_{C}^{\top}\bm{y} \\
         & =\underset{\bm{x}\in {{S}_{I}}}{\mathop{\min }}\,\underset{\bm{\nu }\in {{S}_{D}}}{\mathop{\max }}\,{{\left( \bm{z}-{{\bm{A}}_{I}}\bm{x} \right)}^{\top}}\bm{\nu}+\bm{c}_{I}^{\top}\bm{x}, \label{eq:MILP_VF_dual}
    \end{align}
\end{subequations}
where ${{S}_{D}}=\left\{\bm{\nu}\in{{\mathbb{R}}^{r}}:\bm{A}_{C}^{\top}\bm{\nu}\le{{\bm{c}}_{C}}\right\}$ is the dual feasible region.

Equation (\ref{eq:MILP_VF_dual}) shows that given any integer solution $\bm{x}\in {{S}_{I}}$, $V(\bm{z})$ is the maximum of many affine functions in $\bm{z}$ (thus is a convex polyhedral function); and the full description of $V(\bm{z})$ can be obtained by taking the minimum of $\left|S_I\right|$ convex polyhedral functions. In fact, a subset of $S_{I}$ can be sufficient to characterize $V(\bm{z})$~\cite{hassanzadeh2014on}, because for some integer solutions $\hat{\bm{x}}$, profile $( \bm{z}, \mathop{\max}_{\bm{\nu }\in {{S}_{D}}}\,{{\left( \bm{z}-{{\bm{A}}_{I}}\hat{\bm{x}} \right)}^{\top}\bm{\nu} + \bm{c}^{\top}_I\hat{\bm{x}}} )$ lies on the interior of the epigraph\footnote{The epigraph is ${\rm{epi}}(V) = \left\{(\bm{z}, v) : \bm{z} \in Z, v \in \mathbb{R}, v \ge V(\bm{z})\right\} \subseteq \mathbb{R}^{r+1}$.} of $V(\bm{z})$, so these integer solutions are unessential to describe it. We denote by $S^{\text{min}}_{I}$ the minimal subset of $S_{I}$ required to describe $V(\bm{z})$.

Reference \cite{wolsey2014integer} provides a formal description of this function: the MILP value function is lower semi-continuous, subadditive, and piecewise polyhedral over $Z$. An example to show such properties will be given by Fig.~\ref{fig:SCUC_VF_figure} in Section~\ref{sec:case_2-d_case}.


\subsection{Shapley Value of Cooperative Games}\label{sec:preliminary_Shapley}
Cooperative games deal with the problem of determining a fair and stable division of surplus earned by the coalition, which is formed by rational players. We regard the M-SCUC problem as a \emph{transferable utility} (TU) game, which means that \emph{payments} are allowed between players~\cite{hu2006allocation, zheng2016cooperative, chapman2017cooperative}.
\begin{defination}\label{def:game}
A TU game is given by $\Gamma = \langle \mathcal{A}, v\rangle$ where:
\begin{itemize}
  \item $\mathcal{A}$ is the set of $n$ ($n = \left|\mathcal{A}\right|$) rational players;
  \item $v(\mathcal{C})$ is a characteristic function, $v: 2^{n} \to \mathbb{R}_{+}$ with
$v(\varnothing) = 0$, that maps from each possible coalition $\mathcal{C}\subseteq \mathcal{A}$ to the worth of $\mathcal{C}$.
\end{itemize}
\end{defination}

In the game of M-SCUC, each area incurs some \emph{cost} in producing power, and the areas (players) agree to form a \emph{grand coalition} and work together by coordinately interchange scheduling. Through exchanging power in an efficient way, the total cost should be less than that is achieved by dispatching alone or forming some smaller coalitions. In order for the equality and stability of the grand coalition, the Shapley value is introduced to divide the surplus. The Shapley value is a function that assigns to the characteristic function a $n$-tuple of real numbers, i.e., $\bm{\phi}(v)=\left(\phi_{1}(v), \phi_{2}(v), \ldots, \phi_{n}(v)\right)$. It can be calculated with the belowing formula,
\begin{equation}\label{eqn:Shapley_value}
    \phi_{a}(v)=\sum_{\mathcal{C} \subseteq \mathcal{A} \backslash\{a\}} \frac{|\mathcal{C}| !(|\mathcal{A}|-|\mathcal{C}|-1) !}{|\mathcal{A}| !}(v(\mathcal{C} \cup\{a\})-v(\mathcal{C})).
\end{equation}
In (\ref{eqn:Shapley_value}), the second part $v(\mathcal{C} \cup\{a\})-v(\mathcal{C})$ is the marginal
contribution when player $a$ enters coalition $\mathcal{C}$, and the fraction on the left side is a coefficient indicating the probability that player $a$ enters a coalition of this size.

Shapley value is the unique solution that satisfies the Shapley axioms~\cite{shapley1953value}, i.e., efficiency, symmetry, dummy axiom and additivity.
Moreover, it can be proved that the game of M-SCUC is superadditive and convex~\cite{shapley1971cores}, and hence the Shapley value is individually rational and stable, i.e., no small coalitions (including individuals) could obtain more by acting alone than they can receive from $\bm{\phi}$.

\section{Construction of the SCUC Value Function}\label{sec:construction}
In this section, an algorithm for constructing the set $S^{\text{min}}_{I}$ is presented. We're aware that the authors in~\cite{hassanzadeh2014on} have proposed a cutting plane algorithm, which starts at a singleton in $S_{I}$ and then dynamically generates the set $S^{\text{min}}_{I}$. Yet, the algorithm assumes there exists $\hat{\bm{x}}\in {{S}_{I}}$ so that ${{\bm{A}}_{I}}\hat{\bm{x}}=\bm{z}\ne \bm{0}$, and this is not practical in the SCUC case, e.g., it requires that $\bm{0} = \bm{d}_a-\bm{z}_a \ne \bm{0}$ for constraint (\ref{eqn:M-SCUC_B}). In this paper, we propose a more sophisticated algorithm for the construction of $S^{\text{min}}_{I}$, which is detailed in Algorithm \ref{alg:1}.

\begin{algorithm}
\caption{Construction of the MILP Value Function}
\small
\label{alg:1}
\begin{algorithmic}
\STATE \textbf{Initialize:} Choose an initial right-hand side vector $\bm{z}^0\in Z$, derive an initial integer solution ${\bm{x}}^{0}={\mathop{\arg\min}}_{\bm{x}\in S_{I}({{\bm{z}}^{0}}),\bm{y}}\bm{c}_{I}^{\top}\bm{x}+\bm{c}_{C}^{\top}\bm{y}$, let iteration counter $k=0$, set ${{S}^{k}}=\{ {\bm{x}}^{0} \}$, deviation ${{\Delta }^{k}}=+\infty$ and maximum size of set be $K$.
\WHILE{${{\Delta }^{k}}>0$ and $k < K$}
\STATE $\bullet$ Solve (\ref{eq:global_search}) to obtain a new integer solution ${{\bm{x}}^{k+1}}$
      \begin{equation}
        \indent \indent {{\Delta }^{k}}=\underset{\bm{z}\in Z}\max~\overline V(\bm{z})-V(\bm{z})
        \label{eq:global_search}
      \end{equation}
      where
      \begin{flalign*}
      \indent \indent \overline{V}(\bm{z})=\min_{i=0,...,k} & \bm{c}_{I}^{\top}\bm{x}^{i}+\bm{c}_{C}^{\top}\bm{y}^{i}+M\cdot {{\textbf{1}}^{\top}}\left( {{\bm{s}}^{i}_{+}}+{{\bm{s}}^{i}_{-}} \right) &\\
      \indent \indent \operatorname{s.t.~}& {{\bm{A}}_{I}}\bm{x}^{i}+{{\bm{A}}_{C}}\bm{y}^{i}+\bm{E}{{\bm{s}}^{i}_{+}}-\bm{E}{{\bm{s}}^{i}_{-}}=\bm{z} \\
      \indent \indent & \bm{x}^{i}\in {{S}^{k}},~{{\bm{s}}^{i}_{+}} {\text{~and~}} {{\bm{s}}^{i}_{-}} {\text{nonnegtive}} \\
      \indent \indent & \bm{E}~{\text{an identity matrix}},~M~{\text{a penalty factor.}}
      \end{flalign*}
\STATE $\bullet$ Set ${{S}^{k+1}}\leftarrow{{S}^{k}}\cup \{ {{\bm{x}}^{k+1}} \}$,~$k\leftarrow k+1$.
\ENDWHILE
\STATE \textbf{return} ${{S}^{k}}$
\end{algorithmic}
\end{algorithm}

The main idea of problem (\ref{eq:global_search}) is to find a $\bm{z}$ value under which the current approximation is mostly distinct from the true value function, and generate a relevant ${{\bm{x}}^{k+1}}\in S^{\min}_I\backslash S^{k}$. Problem (\ref{eq:global_search}) can be recast as a mixed integer nonlinear programming (MINLP) below,
\begin{align}
 \nonumber &{{\Delta }^{k}} = \underset{\bm{z}\in Z}\max~\theta - (\bm{c}_{I}^{\top}\bm{x}+\bm{c}_{C}^{\top}\bm{y})  \\
 \nonumber \operatorname{s.t.~} & \bm{\nu}^{i}\in S_D \cap \left[-M,M\right]^r ~i=0,...,k, ~\left(\bm{x},\bm{y} \right)\in S(\bm{z})\\
 &  \theta \le {{\left( \bm{z}-{{\bm{A}}_{I}}{{\bm{x}}^{i}} \right)}^{\top}}{{\bm{\nu}}^{i}}+\bm{c}_{I}^{\top}{{\bm{x}}^{i}} ~~i=0,...,k.
 \label{eq:MINLP}
\end{align}
Problem (\ref{eq:MINLP}) can be solved by solvers like BARON and COUENNE, but it is more efficient to reformulate it as a MILP problem. In this paper, by employing a MILP reformulation technique, namely piecewise McCormick relaxation to the bilinear term ${\bm{z}^{\top}\bm{\nu}^{i}}$, we then approximately solve (\ref{eq:MINLP}) as a MILP problem denoted as ${\underline{{\Delta}}^{k}}$. The reformulation procedure can be found in~\cite{wittmann2013global}, which deals with the multi-parametric MILP problems. The optimal value ${\underline{{\Delta}}^{k}}$ yielded from the the MILP reformulation is an approximation to that of (\ref{eq:MINLP}), however, the sub-optimality will not affect the process of finding new integer solution once ${\underline{{\Delta}}^{k}}$ is positive (a new ${{\bm{x}}^{k+1}}$ can be generated). Ideally, Algorithm \ref{alg:1} will return ${{S}^{k}} = {{S}^{\min}_{I}}$.

In our practice of computation, we set $K$ as a proper number to have Algorithm 1 terminate within an acceptable period of time and return a subset of $S^{\min}_{I}$. In engineering practice, typical historical unit commitment solutions can be added into $S^{k}$ to accelerate the convergence.

\section{Decentralized Solution Method of M-SCUC and Savings Allocation}\label{sec:solution}
\subsection{Solving Decentralized M-SCUC with Value Functions}\label{sec:solution_scheduling}
As the set $S^{\min}_{I}$ is obtained, the characterization of the change of \emph{optimal generation cost} in a full range of interchange level could be identified. With such characterization information of every area ($S^{\min}_{Ia}$), intuitively, the coordinator is able to work out an interchange scheduling that \emph{globally} minimizes the joint cost of the interconnected power system.

Recall that in (\ref{eqn:M-SCUC_obj_new}) the M-SCUC problem has been written as ${\mathop{\min}_{\bm{z}_{a}}}\sum_{a\in \mathsf{\mathcal{A}}}{V_a(\bm{z}_{a})}$, and now we have,
\begin{subequations} \label{eqn:SCUC_VF}
    \begin{align}
        \nonumber {V_a(\bm{z}_{a})} & \overset{(\text{\ref{eq:MILP_VF_dual}})}{=} \underset{\bm{x}_a\in {{S}_{Ia}}}{\mathop{\min }}\,\underset{\bm{\nu }_a\in {{S}_{Da}}}{\mathop{\max }}\,{{\left( \tilde{\bm{z}}_a-{{\bm{A}}_{Ia}}\bm{x}_a \right)}^{\top}}\bm{\nu}_a+\bm{c}_{Ia}^{\top}\bm{x}_a~~~~~~ \\
        & ~=~ \underset{{{\bm{\nu}}^{i}_a},\theta_a}{\max}\, \theta_a \label{eqn:SCUC_VF_LP_obj} \\
        \nonumber & ~~~~~\operatorname{s.t.~} i=0,...,k_a\\
        & ~~~~~~~~~~~~\theta_a \le {{( \tilde{\bm{z}}_a-{{\bm{A}}_{Ia}}{{\bm{x}}^{i}_a} )}^{\top}}{{\bm{\nu}}^{i}_a} + \bm{c}_{Ia}^{\top}{{\bm{x}}^{i}_a} \label{eqn:SCUC_VF_LP_theta} \\
        & ~~~~~~~~~~~~\bm{A}_{Ca}^{\top}\bm{\nu}^{i}_a \le \bm{c}_{Ca}, \label{eqn:SCUC_VF_LP_dual}
    \end{align}
\end{subequations}
where $k_a = |S^{\min}_{Ia}|$. Equation (\ref{eqn:SCUC_VF}) shows that with $S^{\min}_{Ia}$ in hand, the value of $V_a(\bm{z}_{a})$ can be achieved simply by solving a LP problem, i.e., $\max_{{{\bm{\nu}}^{i}_a},\theta_a}\theta_a$. By substituting the LP expression of $V_a(\bm{z}_{a})$ into ${\mathop{\min}_{\bm{z}_{a}}}\sum_{a\in \mathsf{\mathcal{A}}}{V_a(\bm{z}_{a})}$, and retaining the constraints on $\bm{z}_a$, we obtain a min-max problem,
\begin{align}\label{eqn:M-SCUC_VF_minmax}
    \nonumber & {\mathop{\min}_{\bm{z}_{a}}}\mathop{\max}_{{{\bm{\nu}}^{i}_a},\theta_a}\sum_{a\in \mathsf{\mathcal{A}}}\theta_a \\
    \nonumber \operatorname{s.t.~}& \text{(\ref{eqn:M-SCUC_z-range})},\text{(\ref{eqn:M-SCUC_consensus})} \\
    \nonumber & \text{(\ref{eqn:SCUC_VF_LP_theta})} ~~\forall a\in\mathcal{A} ~~:\chi_{a}^{i} \\
    & \text{(\ref{eqn:SCUC_VF_LP_dual})} ~~\forall a\in\mathcal{A} ~~:\bm{\gamma}_{a}^{i}.
\end{align}
%
It can be proved that (\ref{eqn:M-SCUC_VF_minmax}) is equivalent with the MILP below,
\begin{align}\label{eqn:M-SCUC_VF_single-stage_MILP}
    \nonumber & \mathop{\min}_{\bm{z}_{a},\chi_{a}^{i},\bm{\gamma}_{a}^{i}}\sum_{a\in \mathsf{\mathcal{A}}} \sum_{i=1}^{k_a} \chi_{a}^{i}\bm{c}_{Ia}^{\top}\bm{x}_{a}^{i} + \bm{c}_{Ca}^{\top}\bm{\gamma}_{a} \\
    \nonumber \operatorname{s.t.~} & \text{(\ref{eqn:M-SCUC_z-range})},\text{(\ref{eqn:M-SCUC_consensus})} \\
    \nonumber & \sum_{i}^{k_a}{\chi_{a}^{i}(\bar{\bm{z}}_a-\bm{A}_{Ia}\bm{x}_{a}^{i})} + \bm{z}_a - {\bm{A}_{Ca}\bm{\gamma}_{a}} = \bm{0} ~~\forall a\in\mathcal{A} \\
    & \sum_{i}^{k_a}{\chi_{a}^{i}} = 1 ~~\chi_{a}^{i}\in\mathbb{B} ~~\forall a\in\mathcal{A}
\end{align}
where $\bar{\bm{z}}_a$ is a vector obtained by substituting entries of $\bm{z}_a$ in $\tilde{\bm{z}}_a$ with zeros. Problem (\ref{eqn:M-SCUC_VF_single-stage_MILP}) has the following intuitive interpretation: binary variable $\chi_{a}^{i}$ picks up a unit commitment solution from set $S_{a}^{\min}$ for each area, $\bm{z}_{a}$ decides the tie-line injection, while $\bm{\gamma}_{a}$ determines the optimal dispatch of area $a$. Although (\ref{eqn:M-SCUC_VF_single-stage_MILP}) looks like the primal M-SCUC problem~(\ref{eqn:M-SCUC}), solving problem~(\ref{eqn:M-SCUC_VF_single-stage_MILP}) is quite different and it enjoys some attractive properties:
\begin{itemize}
  \item \emph{Decentralization}. In the proposed scheme, each area sets tie-line injections $\bm{z}_{a}$ as variational parameters, and computes a set of unit commitment solutions $S^{\min}_{Ia}$ required to describe its SCUC value function; then the coordinator employs $S^{\min}_{Ia}$ to solve (\ref{eqn:M-SCUC_VF_single-stage_MILP}) and determines the interchange scheduling. The scheme is suitable for multi-area power systems with a hierarchical control architecture.
  \item \emph{Privacy}. When passing data to the coordinator, area $a$ can send $\bm{c}_{Ia}^{\top}\bm{x}_{a}^{i}$ as a scalar $\alpha_{a}^{i}$, and $(\bar{\bm{z}}_a-\bm{A}_{Ia}\bm{x}_{a}^{i})$ as a vector $\bm{\beta}_{a}^{i}$. Hence, some identifiable structures and data of the regional SCUC problem are removed. However, pricing information $\bm{c}_{Ca}$ and network topology information implied in $\bm{A}_{Ca}$ should be exposed. If this is not preferred, we can be avoid it by using the \emph{direct description} of the SCUC value function\footnote{Direct description means describing the optimal objective value by \emph{a set of affine functions purely of $\bm{z}_a$}, that contain no information of the original MILP problem. Direct description is the purpose of multi-parametric MILP~\cite{wittmann2013global}. We will give a demonstration of this with Fig.~\ref{fig:SCUC_VF_figure} latter in Section~\ref{sec:case_2-d_case}.}, instead of the integer solution set description. Since $S^{\min}_{Ia}$ is available now, to obtain the direct description, we only have to solve $| {k_{a}} |$ multi-parametric LPs.
  \item \emph{Computational efficiency}. The number of binary variables, given by $\sum\nolimits_{a\in \mathsf{\mathcal{A}}}{| {k_{a}} |}$, is usually smaller than that in M-SCUC (\ref{eqn:M-SCUC}). Solving problem~(\ref{eqn:M-SCUC_VF_single-stage_MILP}) is always less time-consuming in large-scale problems. This will be demonstrated latter in Section \ref{sec:case_decentralized}.
  \item \emph{Reliability}. Problem~(\ref{eqn:M-SCUC_VF_single-stage_MILP}) is one-shot. Compared with iterative algorithms, solving problem~(\ref{eqn:M-SCUC_VF_single-stage_MILP}) has no convergence problems and durative communication demand.
\end{itemize}

Finally, we have the following Proposition for problem~(\ref{eqn:M-SCUC_VF_single-stage_MILP}):
\begin{proposition} \label{prop:optimality}
If the UC solution set from every regional system is complete, i.e., $k_a = |S_{a}^{k}| = |S_{Ia}^{\min}|~\forall a\in\mathcal{A}$, then the interchange scheduling yielded from (\ref{eqn:M-SCUC_VF_single-stage_MILP}) globally optimizes the M-SCUC problem; if $k_a < |S_{Ia}^{\min}|$ for some $a\in\mathcal{A}$, then the interchange scheduling is sub-optimal.
\end{proposition}

\subsection{Allocating Savings Using the Shapley Value and SCUC Value Functions}\label{sec:solution_allocation}
Traditionally, market clearing methods based on LMPs are adopted to determine the payment of energy trading between areas, e.g., $\mathit{payment}$$=$$\mathit{energy}$$\times$$\mathit{LMP}$, where $\mathit{LMP}$ is the optimal Lagrangian multiplier associated with the row containing a nodal injection parameter. From another point of view, as clarified in Section \ref{sec:preliminary_Shapley}, coordinated interchange scheduling is a cooperative game that leads to surplus, and hence the payments between regional systems can be derived from the Shapley value.

The characteristic function $v(\mathcal{C})$ is essential to calculate the Shapley value. Basically, we have $\mathit{worth}$$=$$\mathit{revenue}$$-$$\mathit{cost}$. Without loss of generality, we assume the revenue of each area to be constant so that it can be neglected. Thus, $v(\mathcal{C})$ can be obtained simply from the generation cost, i.e.,
\begin{align}\label{eqn:characteristic function}
    \nonumber v(\mathcal{C}) = -\mathop{\min}_{\bm{z}_{a}}& \sum_{a\in {\mathcal{C}}} V_a(\bm{z}_a)\\
    \nonumber \operatorname{s.t.~} & \text{(\ref{eqn:M-SCUC_z-range})},\text{(\ref{eqn:M-SCUC_consensus})} \\
    & \bm{z}_a = \bm{0} ~~ \forall a\in \mathcal{A}\backslash\mathcal{C}.
\end{align}
The last constraint in problem~(\ref{eqn:characteristic function}) implies that power can be exchanged within a coalition only. Problem~(\ref{eqn:characteristic function}) can also be reformulated as a MILP problem similar with (\ref{eqn:M-SCUC_VF_single-stage_MILP}).

Finally, as the payoff vector $\bm{\phi}(v)$ is yielded, the payment between areas can be computed according to
\begin{equation} \label{eqn:payment}
  \phi_{a}(v) = -V_a(\bm{z}_a^{\ast}) - \psi_{a},
\end{equation}
where $\bm{z}_a^{\ast}$ is the optimal solution of problem~(\ref{eqn:M-SCUC_VF_single-stage_MILP}), $V_a(\bm{z}_a^{\ast})$ represents the real generation cost of area $a$ acting in the grand coalition, and $\psi_{a}$ is the payment of area $a$ (the \emph{utility} area $a$ should transfer to the others). The interpretation of formula~(\ref{eqn:payment}) is: under the interchange scheduling $\bm{z}_a^{\ast}$, area $a$ has profit $-V_a(\bm{z}_a^{\ast})$, excluding the worth of meeting load demand or the revenue of selling electricity to local consumers; in order to make the payoff \emph{fair}, area $a$ pays $\psi_{a}$ (can be negative) to other areas; ultimately, the net payoff of area $a$ should be equal to the Shapley value $\phi_{a}(v)$.

\section{Case Studies} \label{sec:case}
\begin{figure}[!t]
\centering
\subfloat[]{\includegraphics[width=2.8in]{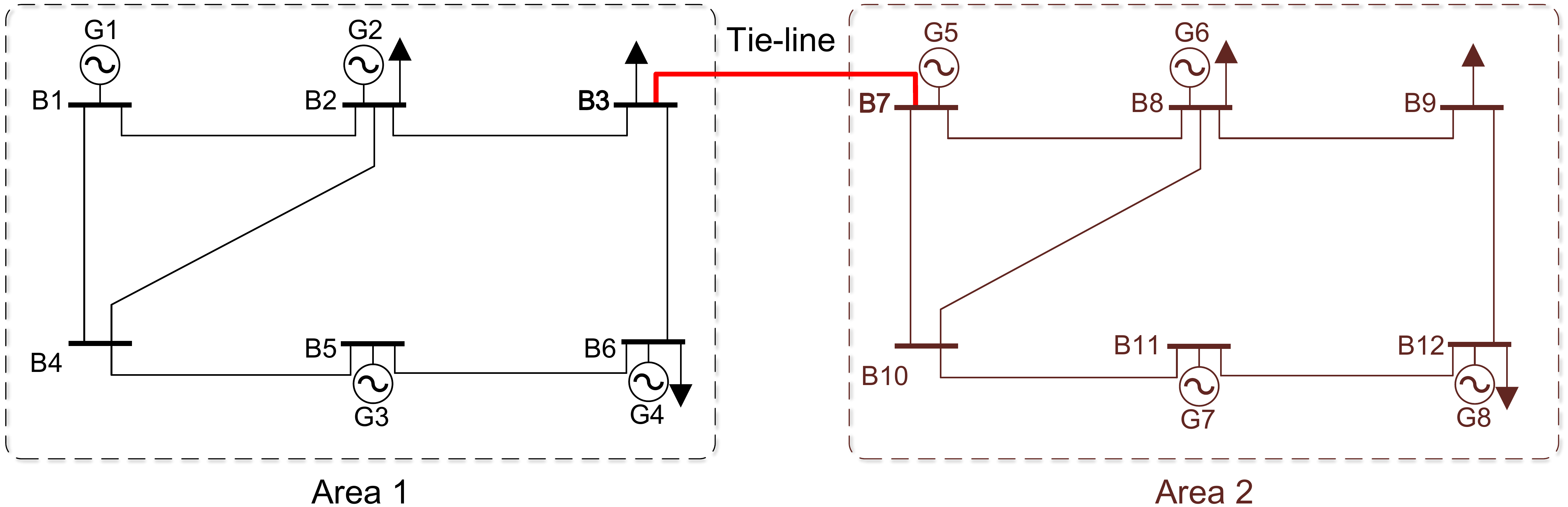}%
\label{fig:diagram_a}}
\hfil
\subfloat[]{\includegraphics[width=1.8in]{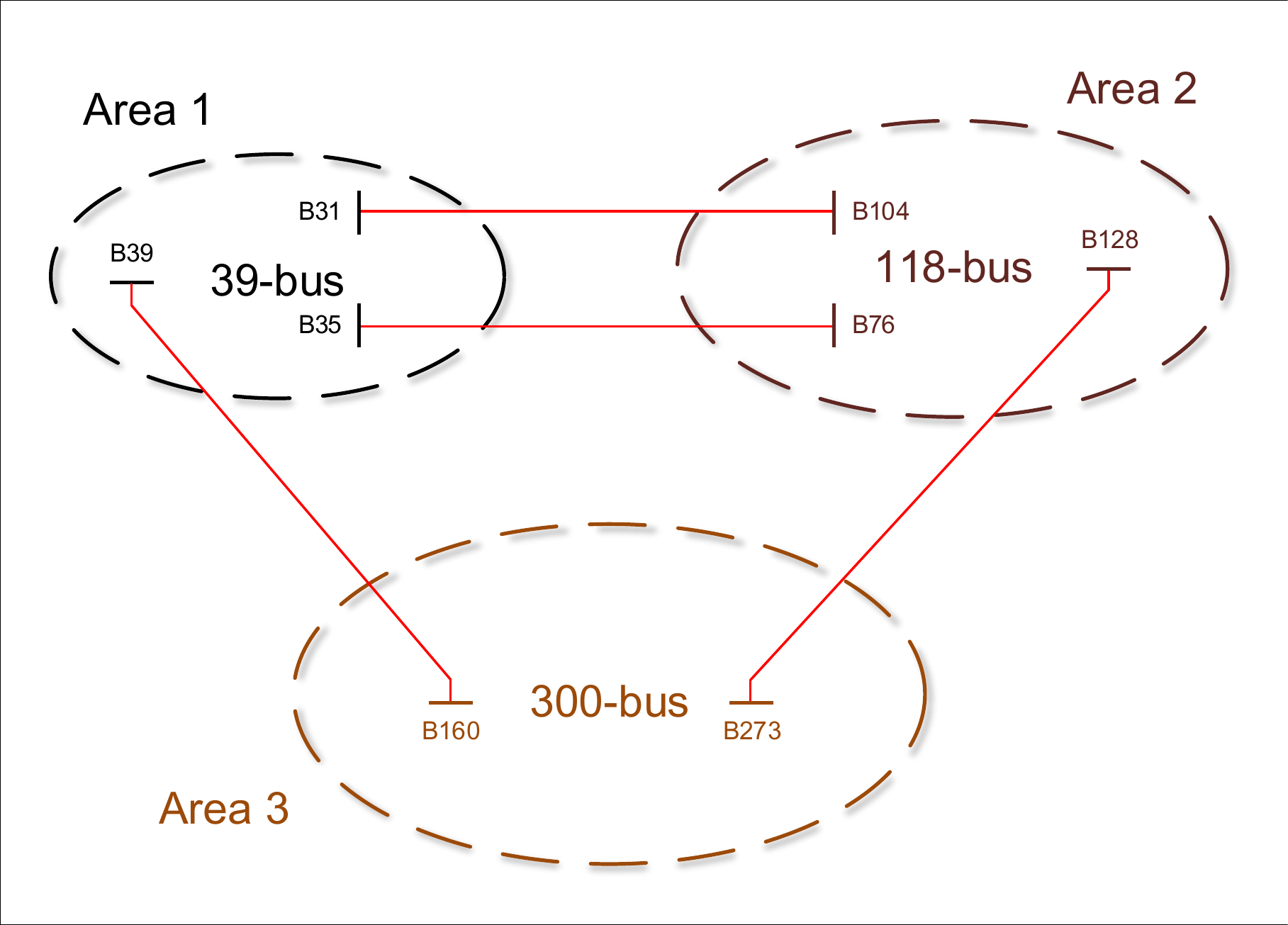}%
\label{fig:diagram_b}}
\caption{Diagrams of the test systems. (a) two-area sytem. (b) three-area sytem.}
\label{fig:diagram}
\end{figure}

In this section, numerical experiments\footnote{MILPs are solved by CPLEX 12.8 on GAMS 26.1; the relative convergence tolerance (optcr) of CPLEX is set as 0. All runs are executed on an Intel i5 CPU machine running at 1.80 GHz with 8 GB of RAM.} are conducted on two multi-area systems\footnote{Available: https://figshare.com/articles/SCUC-VF/7775825/1} to verify the effectiveness and efficiency of the proposed method. Main information of the test systems is summarized in Table~\ref{tab:systems}. The diagrams of them are given in Fig.~\ref{fig:diagram}.
The two-area system comprises two 6-bus systems interconnected by a tie-line from bus 3 to bus 7; the three-area system comprises the 39-bus, 118-bus and 300-bus systems interconnected by four tie-lines.
\begin{table}[t]
\footnotesize
\centering{}%
\caption{Statistics of the Test Systems.\label{tab:systems}}
\begin{tabular}{ccccc}
\hline
System & Composition                             & Buses & Units & Tie-lines \\ \hline
two-area & IEEE 6-, 6-bus                  & 12    & 8     & 1         \\
three-area & IEEE 39-, 118-, 300-bus & 457   & 133   & 4         \\ \hline
\end{tabular}
\end{table}

\subsection{A One Dimensional Case of the SCUC Value Function} \label{sec:case_2-d_case}
As shown in Definition \ref{def:VF}, the SCUC value function is a real-valued function mapping from ${{\mathbb{R}}^{r}}$ to $\mathbb{R}$, where $r$ is the dimension of nodal injections $\bm{z}$. It is hard to depict a high-dimension function. However, we can allow only one nodal injection parameter to be variational, so that the SCUC value function becomes a one dimensional function.

\begin{figure}
  \centering
  \includegraphics[width=3.2in]{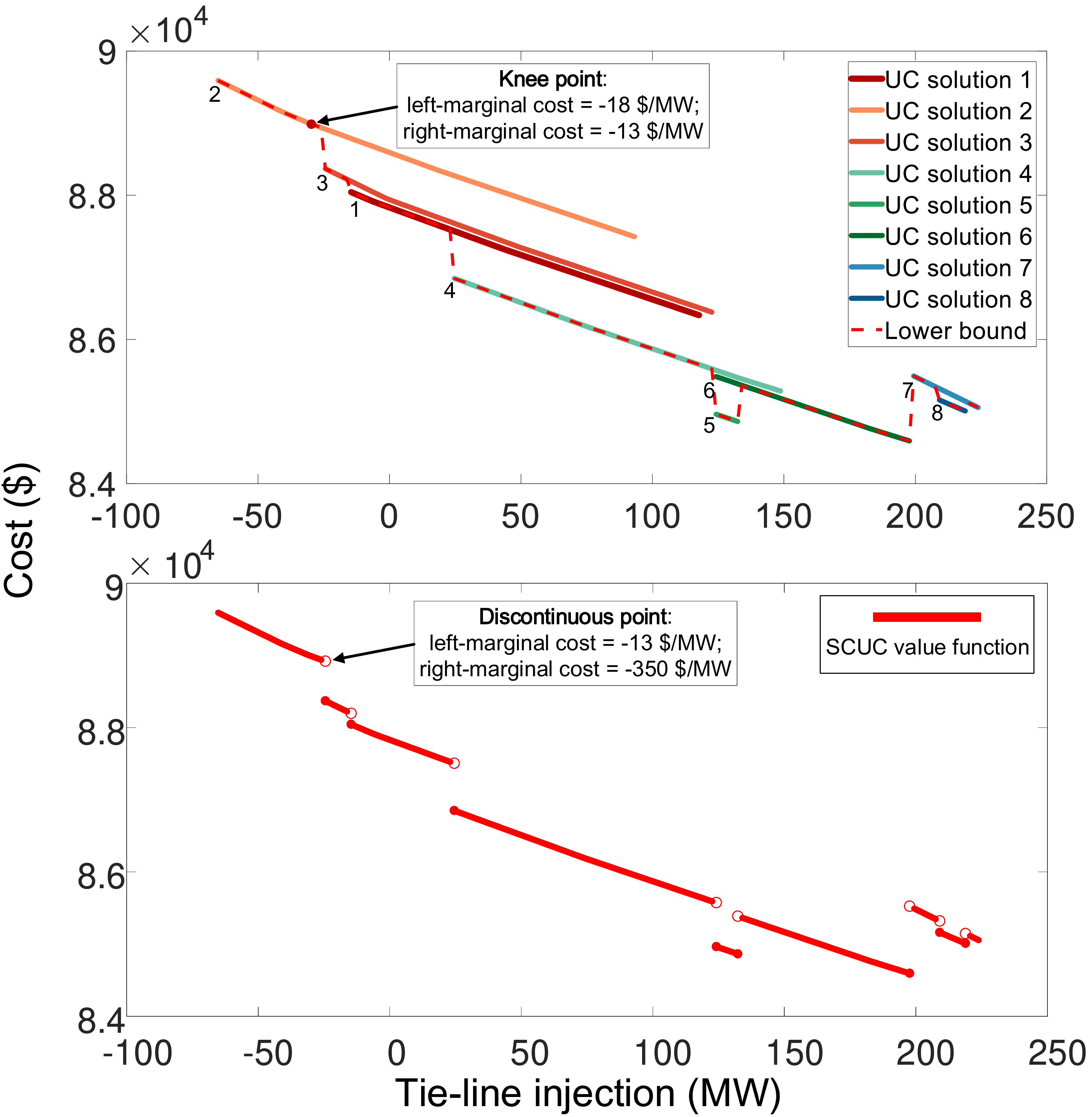}\\
  \caption{SCUC value function of the IEEE 6-bus system with respect to the tie-line injection of bus 3 at one single period, achieved by taking the minimum of eight convex polyhedral functions (LP value functions).}\label{fig:SCUC_VF_figure}
\end{figure}

Specifically, we take the area 1 of the two-area system for example, allow the nodal injection of bus 3 at time 1 to be variational, and use Algorithm~\ref{alg:1} to obtain a set of UC solutions. As shown in the upper piece of Fig.~\ref{fig:SCUC_VF_figure}, for a fixed UC solution, the optimal cost is a piecewise linear function of the tie-line injection (also known as the LP value function). For example, given UC solution 2, the $\mathit{cost}$$-$$\mathit{tie}\text{-}\mathit{line~injection}$ profile has two slopes, namely -18~\$/MW and -13~\$/MW, separated by a knee point.

In this case, eight UC solutions are necessary to describe the SCUC value function. By taking the minimum of these eight LP value functions, we obtain the SCUC value function shown by the lower piece of Fig.~\ref{fig:SCUC_VF_figure}. The value function is composed of \emph{a set of affine functions of $\bm{z}$}, that designates the optimal cost directly in some non-overlapping segments of the tie-line injection. It should be noted that discontinuous points often arise between two different UC solutions (segments).

Figure~\ref{fig:SCUC_VF_figure} can be regarded as a demand/supply curve, and it shows some facts: \emph{i}) if the tie-line injection lies within a segment which has constant slop, the marginal cost is well-defined and the LMP of SCUC is precise; \emph{ii}) if the tie-line injection lies at a knee point or a discontinuous point, the marginal cost isn't well-defined.

At the knee points or the discontinuous points, the left- and right-marginal costs are distinct in general~\cite{stoft2001power}. For example, at the discontinuous point of the SCUC value function depicted in the lower piece of Fig.~\ref{fig:SCUC_VF_figure}, the cost of producing one extra MW more is 350 \$, while that of producing one extra MW less is -13 \$. Due to the ambiguity of marginal cost observed above, investigating other market clearing methods such as game theory based methods may be useful.

\subsection{The Results and Efficiencies of Decentralized M-SCUC} \label{sec:case_decentralized}
In this subsection, we compare the optimality and efficiency between our method (MILP-VF) and the ADMM-AOP method proposed in~\cite{li2015decentralized}. The latter deploys an alternative optimization procedure (AOP) to ensure the convergence of the decentralized algorithm. Specifically, ADMM-AOP iteratively fixes the coupled variables (boundary voltage phase angles) to solve regional SCUC problems separately, and then fixes the binary variables (UC solutions) to solve a decentralized multi-area dispatch problem via ADMM. The centralized optimization method (Centralized) is set as the benchmark.

\begin{figure}[!t]
\centering
\subfloat[]{\includegraphics[width=3.0in]{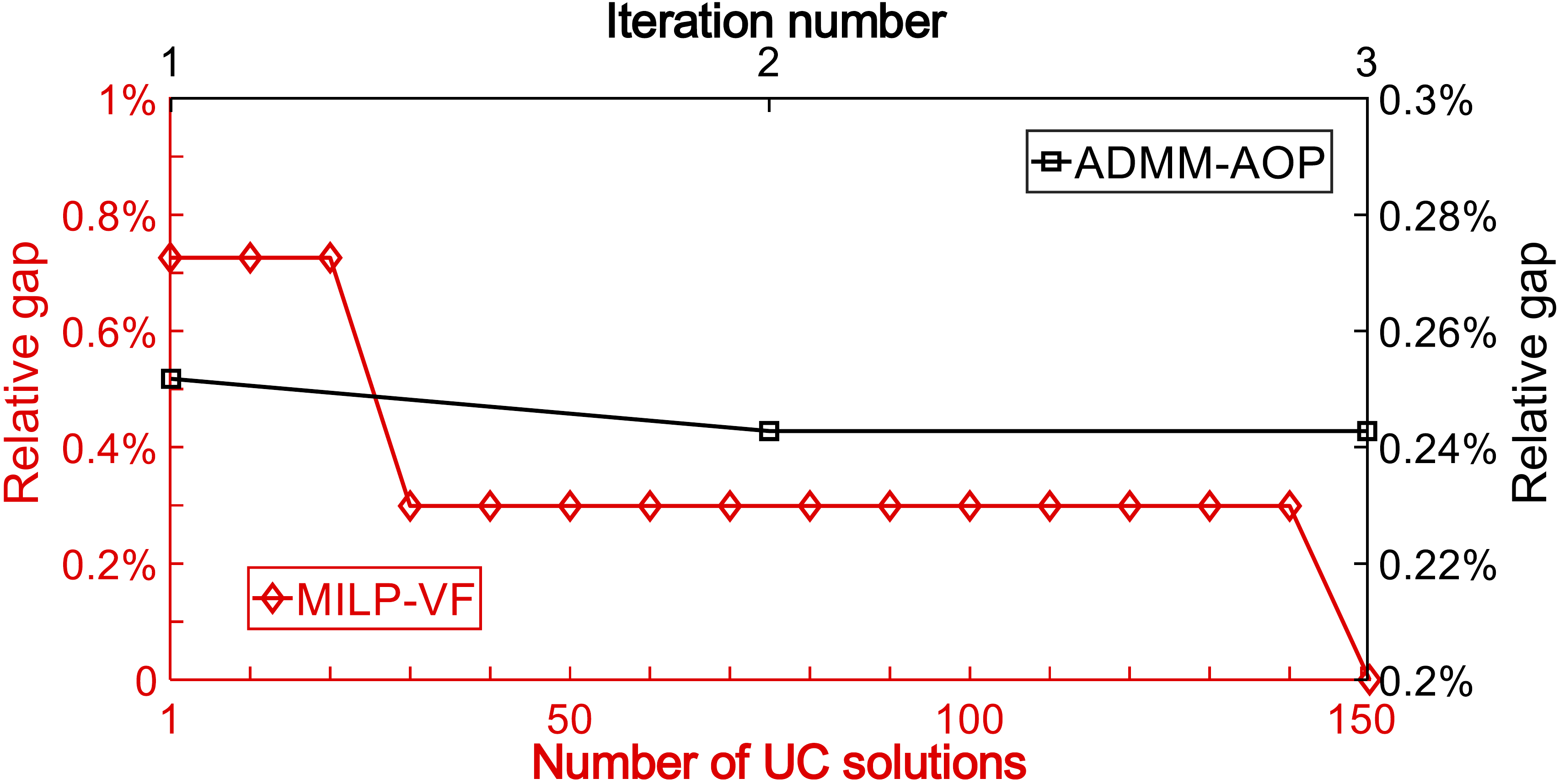}%
\label{1a}}
\hfil
\subfloat[]{\includegraphics[width=3.0in]{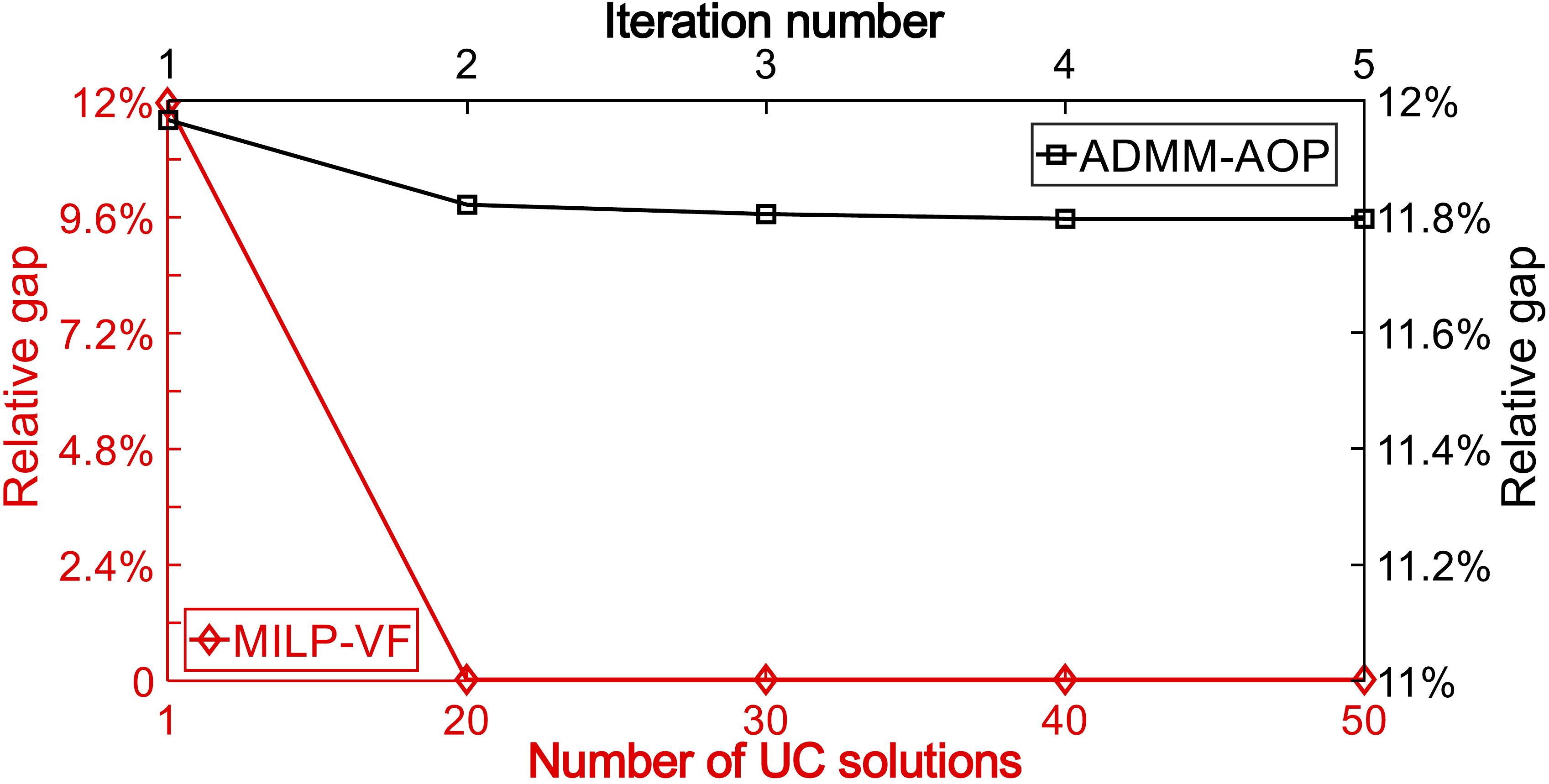}%
\label{1b}}
\caption{The evolutions of relative gaps of MILP-VF method and the convergence profiles of ADMM-AOP method. (a) two-area sytem. (b) three-area sytem.}
\label{fig:Ratio}
\end{figure}

As proved in~\cite{li2015decentralized}, the objective value monotonously decreases with the iteration number of ADMM-AOP; as stated in Section~\ref{sec:solution_scheduling}, MILP-VF is a \emph{non-iterative} method, and its optimality only relies on the size of $S_{a}^{k}$. For both the two-area system and the three-area system, the convergence profiles of ADMM-AOP with respect to the \emph{iteration number} are plotted in the upper axis (black), whereas the evolutions of relative gaps of MILP-VF with respect to the \emph{size of} $S_{a}^{k}$ are plotted in the lower axis (red), all in Fig.~\ref{fig:Ratio}. The non-increasing sequence of relative gaps of MILP-VF indicates that with more identified UC solutions, the interchange scheduling derived from problem (\ref{eqn:M-SCUC_VF_single-stage_MILP}) will be closer to the global optimal solution. The ADMM-AOP method enjoys good convergence property, yet the solutions are easily trapped into local optimums within a few iterations, and hence cannot reach the global optimum.

The gap achieved by MILP-VF in the two-area system is zero, which demonstrates that the MILP-VF method can yield a global interchange solution once the value functions are completely identified. It is also shown in the three-area system that even with a subset of $S^{\text{min}}_{I}$, MILP-VF can obtain a solution very close to that of the centralized method. The relative optimality gaps achieved by ADMM-AOP in two test systems are 0.24\% and 11.80\% respectively. The numerical results are detailed in Table~\ref{tab:cost_gap}, in which the costs when no power is exchanged (Islanded) are also presented.
\begin{table*}[t]
\caption{Comparisons of Joint Cost, Optimality Gap and Runetime in the Two-area and the Three-area Test Systems\label{tab:cost_gap}}
\centering
\small
\begin{tabular}{c|ccc|ccc}
\hline
\multirow{2}{*}{System} & \multicolumn{3}{c|}{Two-area 12-bus} & \multicolumn{3}{c}{Three-area 457-bus} \\
                        & Cost (k\$)  & Gap     & Runtime (s)  & Cost (k\$)  & Gap     & Runtime (s)     \\ \hline
Centralized             & 154.87      & -       & 0.55         & 6,292.26    & -       & 4,111.78        \\
MILP-VF                 & 154.87      & 0.00\%  & 12.85        & 6,293.95    & 0.03\%  & 239.48          \\
ADMM-AOP                & 156.00      & 0.24\%  & 2.69         & 7,034.51    & 11.80\% & 222.77          \\
Islanded                & 68.22+87.83=156.05      & 0.76\%  & 0.48+0.28    & 287.45+965.98+5,945.93=7,199.35    & 14.42\% & 0.99+6.28+11.88 \\ \hline
\end{tabular}
\end{table*}

\begin{figure}
  \centering
  \includegraphics[width=3.0in]{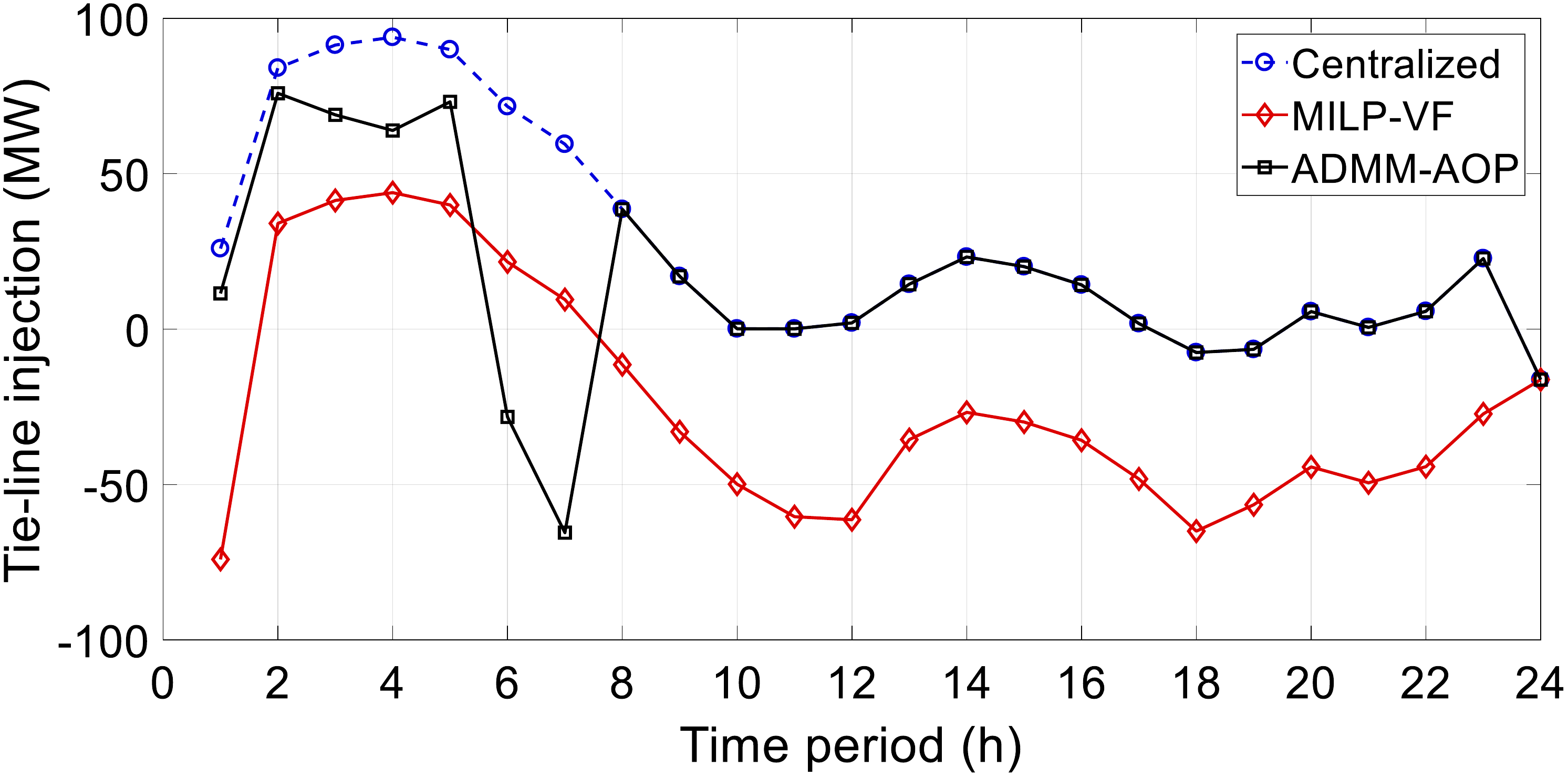}\\
  \caption{Comparison of tie-line power scheduled by the centralized method, the proposed value function based method and the ADMM-AOP method in the two-area system.}\label{fig:pT}
\end{figure}

\begin{figure}
  \centering
  \includegraphics[width=2.8in]{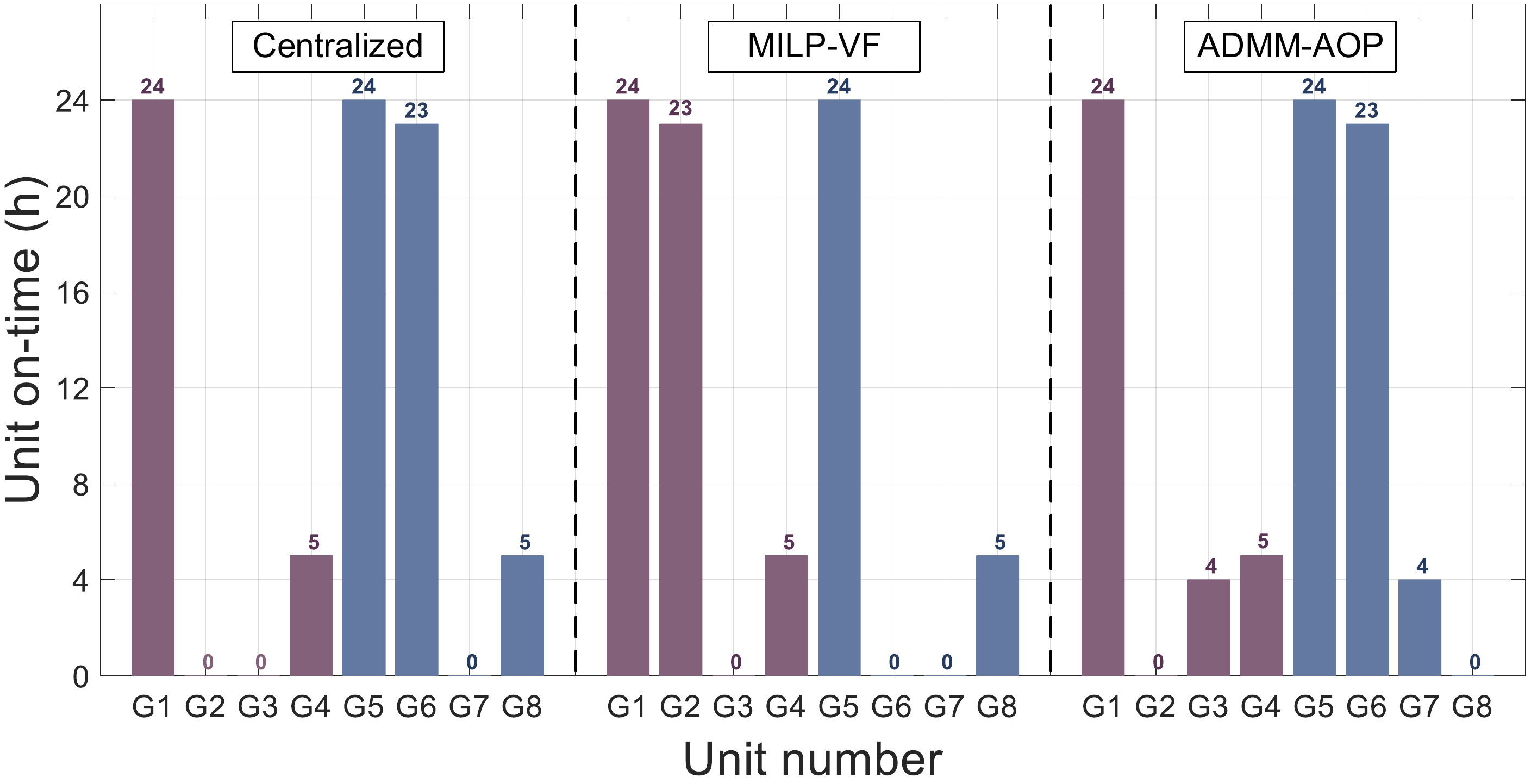}\\
  \caption{Comparison of unit commitment scheduled by the centralized method, the proposed value function based method and the ADMM-AOP method in the two-area system.}\label{fig:unit}
\end{figure}

For the two-area system, we further detail the differences of interchange scheduling and unit commitment achieved by each method. As shown in Fig.~\ref{fig:pT}, the tie-line power of MILP-VF is closer to the optimal one in terms of the shape or tendency, while that of ADMM-AOP is closer to the optimal one in terms of the amount and the direction of power interchange. It is unexpected that the UC solution of MILP-VF shows symmetry with the globally optimal solution, i.e., the unit statues in area 1 of MILP-VF is coincides with those in area 2 of Centralized, and vice versa (see Fig.~\ref{fig:unit}). Note that although area 1 and area 2 have similar system configurations, their load demands are very different. In fact, this case shows that the global optimal solution of MILP is unnecessary unique; it also verifies that the SCUC value functions are highly nonlinear, and hence two very distinct tie-line injection solutions may lead to similar or even identical optimal objective values.

Finally, the computational time of these methods are reported in Table~\ref{tab:cost_gap}. In the two-area case, our value function based method takes more computational time, while the centralized method only takes 0.55 seconds to globally solves the M-SCUC problem. In the three-area case, the runtime of MILP-VF and ADMM-AOP is very close and less than four minutes, whereas the centralized method takes more than one hour. Solving each SCUC problem separately is time-saving in both cases, but always leads to the lowest economic efficiency. Overall, the proposed MILP-VF method can reduce the runtime of solving large-scale M-SCUC problem, while at the same time guarantee a high degree of optimality.

\subsection{Savings Allocation and Comparison with the LMP Method} \label{sec:case_allocation}
Based on the SCUC value functions, the worth of each coalition can be yielded from solving problem~(\ref{eqn:characteristic function}), e.g., $v(\{1\})=-V_1(0)=-68.22$ k\$ (see Table~\ref{tab:cost_gap}).
Then, by substituting the values of $v(\mathcal{C})$ into Equation~(\ref{eqn:Shapley_value}), the payoffs can be obtained. Further, Equation~(\ref{eqn:payment}) is employed to calculate the payments between areas. Take area 1 of the two-area system for example: first, the payoff is
\small
\begin{align*}
  \phi_1(v) = & ~\frac{|\varnothing|!~(|\{1,2\}|-|\varnothing|-1)!}{|\{1,2\}|!}~(v(\{1\})-v(\varnothing)) ~+  \\
  & ~\frac{|\{2\}|!~(|\{1,2\}|-|\{2\}|-1)!}{|\{1,2\}|!}~(v(\{1,2\})-v(\{2\})) \\
  = & ~\frac{1}{2} ((-68.22-0) + (-154.87+87.83))= -67.63,
\end{align*}
\normalsize
and then the payment of area 1 is expressed by $\psi_1 = -77.61  + 67.63 = -9.98$~k\$. Applying the same procedure to area 2 yields 9.98~k\$, which means that area 2 should pay 9.98 k\$ to area 1 (see Table~\ref{tab:payment}).

The payment between area 1 and area 2 calculated via LMPs is slightly different, i.e., 7.72 k\$. We know that LMP is the summation of the marginal costs of power generation and congestion, but it doesn't account for some non-linear or non-convex factors like start-up cost~\cite{hu2006allocation}. On the contrary, the SCUC value functions fully describe the opportunity cost and the marginal contribution of each area, and the payments derived from the Shapley value ensure equity and rationality.
\begin{table}[t]
\caption{Payoffs, Costs and Payments of Each Area in the Two-area System.\label{tab:payment}}
\centering
\small
\begin{tabular}{cc|cc}
\hline
Item                          & Definition              & Area 1      & Area 2      \\ \hline
$\phi_a(v)$                     & payoff (k\$)            & -67.63      & -87.24 \\
$V_a(\bm{z}_a^{\ast})$          & cost (k\$)            & 77.61     & 77.26  \\
$\psi_{a}$                      & payment (k\$)           & -9.98       & 9.98   \\ \hline
\end{tabular}
\end{table}

\begin{figure}
  \centering
  \includegraphics[width=3.2in]{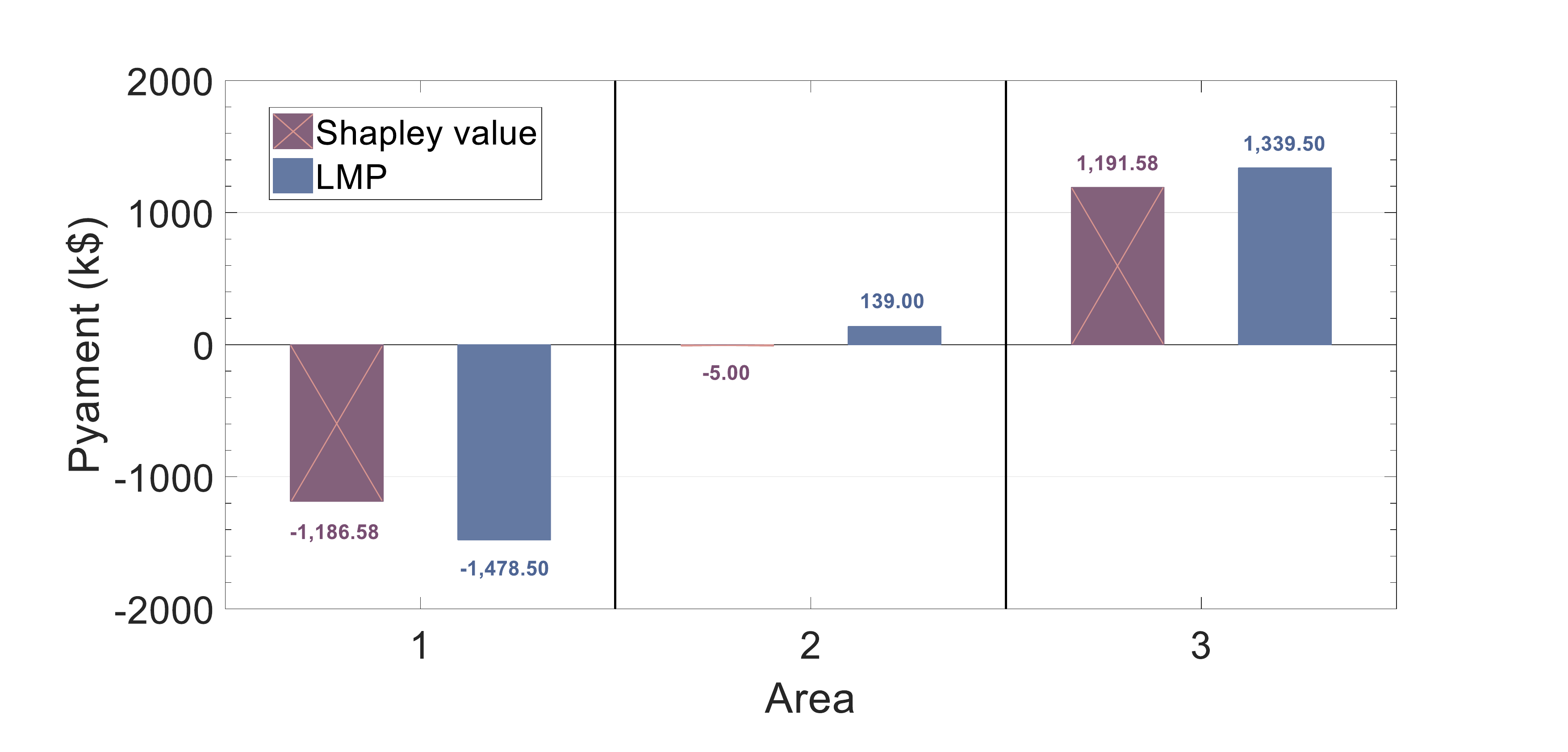}\\
  \caption{Comparison of payments between areas derived from the Shapley value and LPMs.}\label{fig:payment}
\end{figure}

The payments of the three-area system is reported in Fig.~\ref{fig:payment}. It shows that compared with the results of the LMPs based method, area 1 receives less while area 2 and 3 pay less. Moreover, the Shapley value indicates that area 2 can receive 5.00 k\$ even though it has imported 9,699.34 MWh (the average price of electricity is negative). This is counterintuitive, however, it is quite reasonable if one recognizes that without trading with area 2, the total cost of area 1 and area 3 would increase by 171.02 k\$.

Finally, it should be noted that even though LMPs of convex problems are well-defined, and cooperative game theory is also suitable for convex markets, there is no evidence showing that the payments for convex dispatch problems yielded from the LMPs based method and the Shapley value would coincide.

\section{Conclusions}\label{sec:conclusion}
This paper presents a new global solution method for the decentralized multi-area power systems SCUC problem, based on the sets of UC solutions that describe the SCUC value functions. The UC solutions can be generated off-line and repeatedly used, while solving the interchange scheduling problem is computationally efficient, iteration-free and convergence-guaranteed. Even if only a subset of the full set of UC solutions is available, the interchange scheduling yielded from our method is still very close to the globally optimal one. In general, the value function based method proposed in this paper can also be applied to decouple a large-scale MILP problem into tractable sub-problems.

The SCUC value functions constructed here also enable us to investigate how the generation cost is truly affected by the power exchange. They can also be applied directly to derive the Shapley value and the payments of each area. From the perspective of cooperative game theory, the savings allocation scheme obtained from the SCUC value functions and the Shapley value is fair, individually rational and stable. Thus it provides another possible way of market clearing in multi-area power systems apart from the LMPs based methods.

\ifCLASSOPTIONcaptionsoff
  \newpage
\fi

\end{document}